\documentclass[5p]{elsarticle}
\usepackage{graphicx}
\usepackage{multirow}
\usepackage{amsmath,amssymb,amsfonts}
\usepackage{mathtools}
\usepackage{algorithm, algorithmic}
\usepackage{caption}
\usepackage{subcaption}
\usepackage{amssymb}
\usepackage{stfloats}
\DeclareMathOperator*{\argmax}{arg\,max}
\graphicspath{{images/}}

\usepackage{framed} 
\usepackage{multicol} 
\usepackage{nomencl} 
\makenomenclature
\renewcommand*\nompreamble{\begin{multicols}{2}}
\renewcommand*\nompostamble{\end{multicols}}

\usepackage{etoolbox}
\renewcommand\nomgroup[1]{%
  \item[\bfseries
  \ifstrequal{#1}{A}{Abbreviations}{%
  \ifstrequal{#1}{P}{Environment Parameter Symbols}{%
  \ifstrequal{#1}{S}{Reinforcement Learning Symbols}{}}}%
]}

\begin{document}
\begin{frontmatter}

\title{Renewable energy integration and microgrid energy trading using multi-agent deep reinforcement learning
}


\author[label1]{Daniel J. B. Harrold\corref{cor1}}
\address[label1]{School of Computing and Mathematics, Keele University, Keele, United Kingdom}
\address[label2]{Environmental Research and Innovation Department, Sustainable Energy Systems Group, Luxembourg Institute of Science and Technology, Esch-sur-Alzette, Luxembourg}

\cortext[cor1]{Corresponding author.}
\ead{d.j.b.harrold@keele.ac.uk}

\author[label2]{Jun Cao}
\author[label1]{Zhong Fan}


\begin{abstract}
In this paper, multi-agent reinforcement learning is used to control a hybrid energy storage system working collaboratively to reduce the energy costs of a microgrid through maximising the value of renewable energy and trading. The agents must learn to control three different types of energy storage system suited for short, medium, and long-term storage under fluctuating demand, dynamic wholesale energy prices, and unpredictable renewable energy generation. Two case studies are considered: the first looking at how the energy storage systems can better integrate renewable energy generation under dynamic pricing, and the second with how those same agents can be used alongside an aggregator agent to sell energy to self-interested external microgrids looking to reduce their own energy bills. This work found that the centralised learning with decentralised execution of the multi-agent deep deterministic policy gradient and its state-of-the-art variants allowed the multi-agent methods to perform significantly better than the control from a single global agent. It was also found that using separate reward functions in the multi-agent approach performed much better than using a single control agent. Being able to trade with the other microgrids, rather than just selling back to the utility grid, also was found to greatly increase the grid's savings.


\end{abstract}

\begin{keyword}
Actor-Critic Methods, Demand Response, Hybrid Energy Storage Systems, Multi-Agent Systems, Renewable Energy
\end{keyword}

\end{frontmatter}

\begin{table*}[!t]   
	\begin{framed}
		\nomenclature[A]{ANN}{artificial neural network}
		\nomenclature[A]{DDPG}{deep deterministic policy gradients}
		\nomenclature[A]{DQN}{deep Q-networks}
		\nomenclature[A]{ESS}{energy storage system}
		\nomenclature[A]{HESS}{hybrid ESS}
		\nomenclature[A]{LIB}{lithium-ion battery}
		\nomenclature[A]{MAS}{multi-agent system}
		\nomenclature[A]{MDP}{Markov decision process}
		\nomenclature[A]{MGA}{microgrid aggregator}
		\nomenclature[A]{xMG}{external microgrid}
		\nomenclature[A]{PV}{photovoltaic}
		\nomenclature[A]{RL}{reinforcement learning}
		\nomenclature[A]{RES}{renewable energy source}
		\nomenclature[A]{SC}{supercapacitor}
		\nomenclature[A]{TD3}{twin delayed DDPG}
		\nomenclature[A]{VRB}{vanadium redox battery}
		\nomenclature[A]{WT}{wind turbine}
		
		\nomenclature[P, 01]{$x$}{ESS power}
		\nomenclature[P, 02]{X}{power from demand or generation}	
		\nomenclature[P, 03]{$c$}{ESS charge}
		\nomenclature[P, 04]{C}{ESS capacity}
		\nomenclature[P, 05]{P}{energy price}
		\nomenclature[P, 06]{$\eta$}{efficiency}

		\nomenclature[S, 01]{$s$}{state}
		\nomenclature[S, 02]{$a$}{action}
		\nomenclature[S, 03]{$r$}{reward}
		\nomenclature[S, 04]{$\pi$}{policy}
		\nomenclature[S, 08]{$\gamma$}{discount factor}
		\nomenclature[S, 10]{$Q(s,a)$}{critic action-value}
		\nomenclature[S, 11]{$\mu(s)$}{actor action}
		\nomenclature[S, 14]{$\theta$}{critic network weights}
		\nomenclature[S, 15]{$\phi$}{actor network weights}

		
		\printnomenclature
	\end{framed}
\end{table*}

\section{Introduction}
The energy sector is responsible for the overwhelming majority of global greenhouse gas emissions \cite{ritchie_co2_2020}. As the world looks to become more sustainable, a key component of reducing emissions is by moving away from traditional energy generation by increasing the penetration of renewable energy sources (RES) \cite{bogdanov_low-cost_2021}. Although RES can be amongst the most cost effective ways of providing energy \cite{ram_comparative_2018}, the intermittent and unpredictable nature of solar photovoltaic (PV) and wind turbine (WT) generation remains a significant barrier when compared to unsustainable but reliable coal and natural gas power.

Smart energy networks look at tackling this by more intelligently managing the supply and demand of energy. For example, energy storage systems (ESS) may be used to store energy generated from RES when there is a surplus of generation so that it may be used later at peak times. This gives the grid additional stability and security in systems with a large amount of RES \cite{jing_comprehensive_2018}. A hybrid energy storage system (HESS) could also be used in which multiple ESSs are present in the same grid. This can be done to complement the characteristics of the different types, such as pairing a supercapacitor (SC) for short-term storage with a lithium-ion battery (LIB) or fuel cell for longer-term storage \cite{aneke_energy_2016}.

Although this is difficult to achieve at utility grid scales, it can be implemented more easily at smaller scales. Microgrids are local clusters of loads and distributed generation which can either be connected to the main utility grid or act independently as an island. The microgrid may want to use its ESSs for a number of reasons, including RES integration or energy arbitrage, to reduce its dependency on the utility grid \cite{vazquez_energy_2010}. Therefore, a control method is required to be able to learn behaviours for each type of ESS in a complex grid, as well as taking into account fluctuating RES generation and volatile dynamic energy prices.

For this control system, we propose reinforcement learning (RL), a branch of machine learning in which an agent learns to interact with its environment to find an optimal control policy \cite{sutton_reinforcement_2018}. The agent iteratively chooses an action based on its observations of the environment to receive a reward, with the goal to maximise its total future reward. In addition, artificial neural networks (ANN) can be used to combine RL with deep learning for solutions to more complex problems. Model-free RL algorithms are incredibly versatile and easy to implement as they can be used in situations where there is little information or data available on the environment, as the agent learns from its own experiences through trial-and-error.

In this energy network context, the problem can also be considered as a multi-agent system (MAS) in which multiple intelligent decision-making agents must cooperate, coordinate, and negotiate with each other to achieve their individual goals \cite{wooldridge_introduction_2009}. In theory, RL should be a suitable learning mechanism for these agents as it too is used for sequential decision making. However, there are a number of fundamental challenges when implementing RL for a MAS such as a non-stationary environment, communication between agents, and assigning appropriate rewards for agent learning \cite{busoniu_multi-agent_2010}.

Therefore, there is a dilemma if a microgrid energy system should be managed by a single global controller or by multiple distributed agents in a MAS. In this paper, this problem is investigated using two case studies: the first controlling RES to maximise the utilisation and value of RES under dynamic pricing to reduce the grid's energy bill, and the second building on the first using an aggregator trading with other neighbouring microgrids looking to reduce their own bills.

\subsection{Related Literature}
There is a lot of research into the use of RL for ESS control to maximise RES \cite{perera_applications_2021}, and research into the field has increased substantially over the last decade \cite{vazquez-canteli_reinforcement_2019}. Kuznetsova et al. \cite{kuznetsova_reinforcement_2013} where the first authors to use RL to control an ESS in the presence of RES, using a battery to maximise the utilisation of WT generation. However, the algorithm used is considered outdated by current standards of deep RL approaches which were developed several years after.

There is also work on the use of RL for HESS, but it is generally less explored and typically uses a single control agent to manage the multiple ESSs. Francois-Lavet et al. \cite{francois-lavet_deep_2016} used deep RL to control both a LIB and a hydrogen fuel cell to maximise RES utilisation in a solar microgrid. Qiu et al. \cite{qiu_heterogeneous_2016} used RL to control both a lead-acid and a vanadium redox battery (VRB) to reduce grid power losses, again in a solar microgrid. However, as both of these papers only consider two different ESSs, the scalability of these approaches when used in larger microgrids with more ESSs or other types of agents is unclear. Zhang et al. \cite{zhang_data-driven_2021} used a more advanced deep RL algorithm for the control of a microgrid completed with a HESS and both PV and WT generation. The authors test a variety of different single-agent approaches for control of the entire grid, but also suggest exploring the use of a MAS as a potential future research avenue.

There are alternative model-based methods for ESS control. Prodan et al. \cite{prodan_model_2014} compared the use of a model predictive controller to the research of Kuznetsova et al. \cite{kuznetsova_reinforcement_2013} and note better performance but requires significantly more computation. Li et al. \cite{li_improved_2021} use distributed coordinated power control for voltage and frequency control using a battery, SC, and fuel cell in a combined AC/DC microgrid. However, these methods rely on having a complete model available of the environment and their performance is only as accurate as the model itself. In contrast, a RL algorithm could be trained on a digital model but would continue to learn and improve when applied to and evaluated on a real system.

Multi-agent RL has also been used for HESS and larger microgrids that would be difficult to control using a single agent. Foruzan et al. \cite{foruzan_reinforcement_2018} uses RL to control multiple agents in a microgrid including an ESS, RES, a generator, and self-interested customers each with their own reward function. However, the algorithm used is fairly basic and restricts control to a discrete state and action space. Kofinas et al. \cite{kofinas_fuzzy_2018} used the same algorithm but adapted with fuzzy logic to adapt the method for continuous state and action spaces in a MAS. However, more advanced methods have been developed since that can achieve the same thing without the fuzzy logic. Mbuwir et al. \cite{mbuwir_reinforcement_2019} uses multi-agent deep RL to control a battery and heat pump, but the microgrid is much smaller with fewer agents than the two aforementioned works.

To the best of our knowledge, there is no research into multi-agent RL using continuous control for agents. In addition, there is no literature analysing the use of single-agent versus multi-agent RL for RES integration or energy trading in microgrids. 

\subsection{Contributions}
The contributions of this paper are as follows:

\begin{itemize}
\item Presents a novel use of the reinforcement learning algorithm deep deterministic policy gradients (DDPG) and its state-of-the-art variants distributional DDPG (D3PG) and twin delayed DDPG (TD3), as well as their multi-agent equivalents, for continuous control of the ESSs and trading agents.

\item Presents two case studies for the use of RL for energy arbitrage and RES integration in microgrids: the first purely looking at maximising the value of local RES generation to reduce energy costs, and the second by connecting the primary microgrid to other neighbouring microgrids with which it can sell energy to.

\item Examines the use of centralised global microgrid control and distributed multi-agent control, analysing if it is better for one agent to observe and manage the whole environment or have multiple agents responsible for their own individual microgrid component. 

\item Examines the use of deep RL algorithms for continuous control of a HESS with a SC suited for short-term storage, a LIB for medium-term storage, and a VRB for long term storage. 
\end{itemize}

\subsection{Structure}
The rest of this paper is organised into the microgrid environment description in Section \ref{sec:SEN}, the background into RL and the specific algorithms in Section \ref{sec:RL}, outline of the two case studies in Section \ref{sec:methods}, results and discussion in Section \ref{sec:results}, with final conclusions made in Section \ref{sec:conclusion}.
\section{Microgrid Description}
\label{sec:SEN}

\begin{figure*}[!t] 
	\centering
	\includegraphics[width=.80\textwidth]{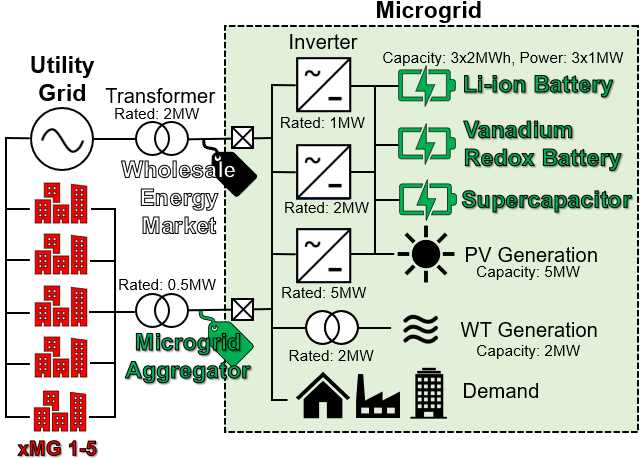}
	\caption{Basic schematic of the microgrid.}
	\label{fig:grid}
\end{figure*}

Microgrids are localised energy networks with their own demand and RES that are able to trade energy with the main utility grid or operate as an independent island \cite{katiraei_microgrids_2008}. This section will describe the primary microgrid used in the future case studies and how it is modelled as a Markov decision process (MDP) for RL. The first case will look solely at using the HESS to maximise the utilisation of the RES to reduce the grids energy bills, and the second will consider using an aggregator to sell energy to neighbouring external microgrids (xMG) competing against each other to reduce their own energy costs. 

\subsection{Environment Description}
The primary microgrid is fitted with a HESS as well as both PV and WT generation. It is connected to the main utility grid that sets dynamic energy prices from which the microgrid can import energy from, or sell back to at a fixed feed-in tariff. AC and DC power lines are present connected via inverters, as well as transformers between the primary microgrid and both the utility grid and WT. In addition, the primary microgrid is also connected to five xMGs via another transformer, with those xMGs are connected themselves to the main utility grid but not the wholesale energy market. The basic schematic of the environment is shown in Figure \ref{fig:grid}.

\subsubsection{Demand and Pricing}
The demand data collected at Keele University Campus ranges from 00:00 January 1st 2014 up to and including 23:00 December 31st 2017. The raw data is separated into different residential, industrial, and commercial sites as well as readings for key individual buildings. The demand used for this simulation is from the three main incomer substations into the campus with the half-hourly readings summed to match the frequency of the hourly weather data. All loads are AC.

The microgrid operates under a dynamic energy pricing scheme where the prices $\text{P}^\text{grid}$ are set by a real-time energy trading market covering the UK \cite{nord_pool_historical_2020}. However, there is also a set maximum price $\text{P}_\text{max} = \text{£}144\text{/MWh}$ as this is the average unit rate for electricity for the UK \cite{uk_power_compare_2021} so will instead buy from the utility grid at that flat price if the market price exceeds this.

The microgrid can also sell energy back to the utility grid at a fixed feed-in tariff. However, the feed-in tariff in the UK varies greatly based on the type of energy, the size of the system, and the installation date \cite{govuk_feed-tariffs_2021}. Therefore, a fixed rate of $\text{P}_\text{min} = \text{£}16\text{/MWh}$ is assumed for all energy sold back to the grid. This is equal to the tariff for a WT system selling between $0.1\text{MW}$ and $1.5\text{MW}$ installed after 1st January 2019, and is also extremely similar to the $\text{£}15.9\text{/MWh}$ feed-in tariff of a $0.25\text{MW}$ and $1\text{MW}$ PV system installed at the same time \cite{ofgem_feed-tariff_2021}.

\subsubsection{Renewable Energy Sources}
The output for the renewable generation is modelled using weather data collected at the Keele University weather station. The WT's output $\text{X}^{\text{WT}}$ uses power curve modelling with wind speed $v$:

\begin{equation}
\text{X}^{\text{WT}} = 
\begin{cases}
	0 & v < v_{\text{ci}} \text{ or } v > v_{\text{co}}\\
	\frac{1}{2}\rho \pi r^2 c_{p}v^{3} & v_{\text{ci}} \leqslant v < v_{r} \\
	\text{X}_{\text{rated}}^{\text{WT}} & v_{r} \leqslant v \leqslant v_{\text{co}}
\end{cases}
\end{equation}

where $v_{ci}=3\text{ms}^{-1}$, $v_{r}=12\text{ms}^{-1}$, and $v_{co}=25\text{ms}^{-1}$ are the cut-in, rated, and cut-out wind-speeds respectively. These values are comparable to a 1MW-rated turbine with a blade radius of $r=30\text{m}$ and a power coefficient $c_p=0.4$ \cite{carrillo_review_2013}. This microgrid considers two of these turbines for a combined maximum wind capacity of $2\text{MW}$. The same power curve approach is used by both Kuznetsova et al. \cite{kuznetsova_reinforcement_2013} and Zhang et al. \cite{zhang_data-driven_2021} for their RES output. The solar output is calculated using hourly solar radiation data and scaled to model a solar farm with a maximum capacity of 5MW. The output of the WT is AC while the output of the PV is DC.

\subsubsection{ESS}
\begin{table*}[t]
\centering
\caption{ESS Properties}
\begin{tabular}{|l|lllllll|}
\hline
ESS & Capacity $\text{C}_\text{max}$ & Power $\text{X}_\text{max}$ & $\eta_\text{SDC}$ & $\eta_\text{RTE}$ & Capacity Cost & Lifecycles & $\text{P}_{\text{CPC}}$ \\
\hline
LIB & 2MWh & 1MW & 99.99\% & 95\% & £100/kWh & 5k & £40 \\
VRB & 2MWh & 1MW & 100\% & 80\% & £200/kWh & 10k & £40 \\
SC & 2MWh & 1MW & 99\% & 95\% & £300/kWh & 100k & £6 \\
\hline
\end{tabular}
\label{tab:ess}
\end{table*}
The primary microgrid consists of a HESS with three different types of ESS: a LIB, a VRB, and a SC. Each ESS is charged and discharged at the same time. The parameters used for the modelling of each type are given in Table \ref{tab:ess} with their maximum capacity $\text{C}_{\text{max}}$, maximum power $\text{X}_{\text{max}}$, self-discharge efficiency (SDC) per hour $\eta_\text{SDC}$, round-trip efficiency (RTE) $\eta_\text{RTE}$, capacity cost per kWh, number of lifecycles, and capacity cost per cycle $\text{P}_{\text{CPC}}$. Some general observations of each are:

\begin{itemize}
\item LIB is cheapest per kWh with a low SDC and a good RTE but has the fewest total lifecycles making it ideal for medium-term storage.

\item VRB has a negligible SDC but poor RTE (around 10\% for each charge or discharge) making it suited for only medium to long-term storage.

\item SC has by far the largest number lifecycles making it by far the cheapest to operate but with a poor SDC (24\% per day) so is ideal for only short-term storage.
\end{itemize}

The charge of each ESS is bound between 0 and their maximum capacity $\text{C}_{\text{max}}=2\text{MWh}$, with the power $x_t$ the ESS can charge or discharge each step bound between 0 and their maximum power $\text{X}_{\text{max}}=1\text{MW}$. The charge of each ESS $c_t$ each step is calculated by:

\begin{equation}
c_t = x_t\sqrt{\eta_{\text{RTE}}} + c_{t-1} \eta_{\text{SDC}}
\end{equation}

The operation CPC of each ESS is also considered so that the agents optimally use the different types. This is calculated using their capacity, capacity cost, and number of lifecycles which is then used as a function in the reward $R^\text{CPC}$:
\begin{equation}\label{eq:rop}
R^\text{CPC}_t = 0.5 \text{P}_{\text{CPC}} \left( \frac{c_t - c_{t-1}}{\text{C}_\text{max}} \right) ^2
\end{equation}

A cycle is considered as the battery charging from empty to full and then discharging until empty, or vice versa. Therefore, $\text{P}_{\text{CPC}}$ is halved as each step is only considered as a half cycle because the ESS cannot both charge and discharge in the same step.

\subsubsection{Transformers and Inverters}
\begin{figure}[!t] 
	\centering
	\includegraphics[width=.45\textwidth]{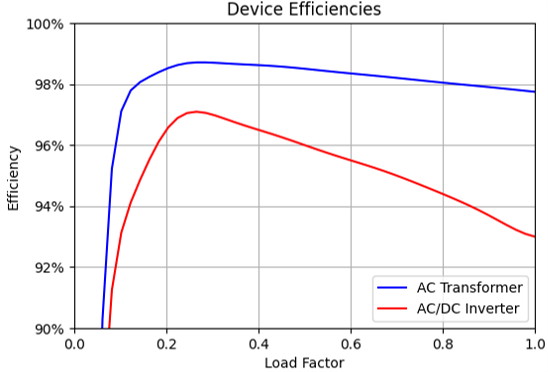}
	\caption{Inverter and Transformer Efficiency Profiles.}
	\label{fig:invtra}
\end{figure}

The primary microgrid is a hybrid AC/DC network with inverters between the two power lines and transformers to step-down the voltage to the microgrid. The efficiency profiles are nonlinear and given as a function of the load factor, which is given as the current load divided by the rated load. The rated loads of each of the transformers and inverters can be found in Figure \ref{fig:grid} with the efficiency profiles modelled using a polynomial estimation, shown in Figure \ref{fig:invtra}. Although the real efficiency of these devices at low load factor would approach 0\%, the minimum efficiency is capped at 10\% to prevent divisions by 0 during environment calculations. 

The HESS and PV generation are connected to 1MW, 2MW, and 5MW rated inverters as well as to each other. When energy is to pass between the AC and DC lines, the grid will automatically use whichever inverter or combination of inverters that would result in the lowest power loss. Mbuwir et al. \cite{mbuwir_reinforcement_2019} used a similar efficiency profile for inverters \cite{driesse_beyond_2008} for RL in a microgrid with solar energy.

\subsection{Microgrid Aggregator (MGA)}
\begin{algorithm}[t] 
\caption{MGA and xMG Bidding Step}
\begin{algorithmic}[1]

\STATE Input MGA selling volume $\text{X}^{\text{MGA}}$
\STATE Input MGA reserve price $\text{P}^{\text{MGA}}$
\STATE Input xMG bidding volumes \\ $x^{\text{xMG}} = \left[ x^{\text{xMG1}}, x^{\text{xMG2}}, x^{\text{xMG3}}, x^{\text{xMG4}}, x^{\text{xMG5}} \right]$
\STATE Input xMG bidding prices  \\ $p^{\text{xMG}} = \left[ p^{\text{xMG1}}, p^{\text{xMG2}}, p^{\text{xMG3}}, p^{\text{xMG4}}, p^{\text{xMG5}} \right]$
\STATE Initialise reward $R_{\text{MGA}} = 0$

\WHILE{$\text{X}^{\text{MGA}} > 0$ and $x^{\text{xMG}}_\text{any} > 0$}
	\STATE Highest bidding xMG index $i = \argmax p^{\text{xMG}}$
	\IF{$p^{\text{xMG}}_i > \text{P}^{\text{MGA}}$}
		\STATE Available bid volume $x^{\text{bid}} = \min[\text{X}^{\text{MGA}}, x^{\text{xMG}}_i]$
		\STATE Update reward $R_{\text{MGA}} \leftarrow R_{\text{MGA}} + 0.8 p^{\text{xMG}}_i x^{\text{bid}}$
		\STATE Update selling volume $\text{X}^{\text{MGA}} \leftarrow \text{X}^{\text{MGA}} - x^{\text{bid}} $
	\ENDIF
	\STATE Set $x^{\text{xMG}}_i \leftarrow 0$
	\STATE Set $p^{\text{xMG}}_i \leftarrow 0$
\ENDWHILE
\STATE Sell remainder $R_{\text{MGA}} \leftarrow R_{\text{MGA}} + \text{P}_{\text{min}} \text{X}^{\text{MGA}}$

\end{algorithmic}
\label{alg:bidding}
\end{algorithm}
In the second case study, the aggregator agent controls the amount and price of energy sold by the main microgrid at the next step. The MGA selects the total volume the grid will sell $\text{X}^{\text{MGA}}$ for the xMGs to bid for. The MGA also selects a reserve price $\text{P}^{\text{MGA}}$ between $\text{P}_\text{min}$ and $\text{P}_\text{max}$ which needs to be met in order to sell so that the MGA has some control over the xMG market. 

During the bidding phase of each step, the aggregator will sell as much available energy as available to the highest xMG bidder up to the volume it has requested, assuming the xMG bid price exceeds the MGA reserve price. The MGA will then move to the next highest bidder until all available energy for selling has gone, with any excess energy not sold given back to the utility grid at $\text{P}_\text{min}$. This is written in pseudocode in Algorithm \ref{alg:bidding}.

\subsection{External Microgrids (xMG)}
The xMG are small microgrids with their own demand, but no ESS or RES. Their demand is equal to one-twentieth of the primary microgrid's demand $\text{X}^{\text{xMG,D}}$ plus a small noise function:

\begin{equation}
\text{X}^{\text{xMG, D}}_t = \left[ 0.05\text{X}^{\text{D}}_t + \mathcal{N}(0, 0.01) \right] ^{0.25\text{X}^{\text{D}}_t}_{0.01\text{X}^{\text{D}}_t}
\end{equation}

The xMG bids for a volume of energy $x^{\text{xMG}}$ at a price $p^{\text{xMG}}$ between $\text{P}_\text{min}$ and $\text{P}_\text{max}$. Any demand that is not met by the volume it receives from the MGA is made up by buying from the utility grid at $\text{P}_\text{max}$. Any energy bought over the xMG demand is given back to the utility grid at $\text{P}_\text{min}$.

\subsection{Markov Decision Process}
For RL to operate the microgrid environment, the control process must be modelled as an MDP. This means the control process must satisfy the Markov property and can be generalised to a state-space $\mathcal{S}$, action-space $\mathcal{A}$, reward function $\mathcal{R}$, and a state transition function. The rest of this section will explain the states, actions, and rewards for the case studies.

\subsubsection{States}
The ESS storage system agents observe the charge of all of the ESSs, the grid demand, wholesale energy price, WT and PV generation, as well as the current hour in the day and the week. The agents also receive predicted values of the demand, price, and RES from ANNs using regression with weather data as in previous work \cite{harrold_data-driven_2022}.

In the second case study, the ESS agents also observe the aggregator sell volume and the reserve price decided by the MGA agent in the previous step. The xMG agents also observe those values along with the same hourly information as the ESSs, as well as their own demand. The demands of each of the microgrids, including the ESS charge and RES generation of the main microgrid, are not shared with any of the other agents in the interest of privacy.

\subsubsection{Actions}
The actions of the agents control the environment at each step. In both case studies, the ESS agents select an action which corresponds to how much it will charge or discharge in the current step, between a maximum charging power $\text{X}_{\text{max}}$ to a maximum discharging power $-\text{X}_{\text{max}}$. The global controller agent will output three actions to control all of the HESS together, whereas the individual controllers in the multi-agent setting select one action each for their corresponding ESS.

In the second case study, the MGA decides the sell volume of energy and the reserve price for the xMG at the next step. These values are then observed by all agents at the next step. In this case, the global controller will output five actions instead of three, whereas the individual MGA agent will operate as its own entity outputting two actions in the multi-agent setting. Each xMG agents decide two actions: the buying volume of energy and the bid price of the current step.

For DDPG, the primary algorithm used in this paper, the actions are selected using an actor network for each agent which observes only the relevant state observations for that particular agent. However, the agents' performance is evaluated using a critic network for each agent which receives those same observations as well as the actions taken by all agents.

\subsubsection{Rewards}
The aim of the agents is to reduce the energy costs of their respective microgrid. For the primary microgrid, this is done by calculating the energy imported to or exported from the grid. First, the net demand of the DC line $\text{X}^{\text{dc}}$ is considered:

\begin{equation}
\text{X}^{\text{dc}}_t = x^{\text{LIB}}_t\eta^{\text{LIB}}_{\text{RTE}} + x^{\text{VRB}}_t\eta^{\text{VRB}}_{\text{RTE}} + x^{\text{SC}}_t\eta^{\text{SC}}_{\text{RTE}} - \text{X}^{\text{PV}}_t
\end{equation}

This is then used to calculate the amount of energy imported from or exported to the utility grid:

\begin{equation}
\text{X}^{\text{in}}_t = (\text{X}^{\text{D}}_t + \text{X}^{\text{dc}}_t\eta^{\text{inv}}_{\text{[1,2,5]MW}} - \text{X}^{\text{WT}}_t\eta^{\text{tra}}_{\text{2MW}})\eta^{\text{tra}}_{\text{2MW}}
\end{equation}

The reward for the cost of the energy the primary microgrid has bought or sold to the utility grid can then be calculated:

\begin{equation}
R^{\text{in}}_t = - \text{X}^{\text{in}}_t \text{P}^{\text{grid}}_t
\end{equation}

The ESS agents are punished to promote positive behaviour, such as by including the ESS operation cost $R^\text{CPC}$ calculated earlier in Equation \ref{eq:rop}. The agent is punished again if the action selected would make the theoretical ESS charge $\dot{c}$ after the action is executed exceed the capacity boundaries of the ESS:

\begin{equation}
R^{\text{cap}}_t = 
\begin{cases}
	\text{P}_{\text{max}}(\dot{c}_t)^2 / \text{X}_{\text{max}} & \dot{c}_t < 0 \\
	0 & 0 \leq \dot{c}_t \leq \text{C}_\text{max} \\
	\text{P}_{\text{max}}(\dot{c}_t - \text{C}_\text{max})^2 / \text{X}_{\text{max}} & \dot{c}_t > \text{C}_\text{max}
\end{cases}
\end{equation}

The agents are also punished based on the amount of energy lost each step through self-discharge:

\begin{equation}
R^{\text{SDC}}_t = \text{P}_{\text{max}} \left( \frac{c_t}{\text{C}_\text{max}} \right)\eta_\text{SDC}
\end{equation}

Including the ESS operation cost $R^\text{CPC}$ from Equation \ref{eq:rop} and the reward for the MGA bidding $R^\text{MGA}$ from Algorithm \ref{alg:bidding}, the sum of the reward terms $R^{\text{sum}}$ is calculated as:

\begin{equation}
R^{\text{sum}} = R^{\text{in}} + R^\text{MGA} - R^{\text{CPC}} - R^{\text{SDC}} - R^{\text{cap}} - R^{\text{base}}
\end{equation}

This value is then multiplied by 0.01 and the total number of agents $N_{\text{agents}}$ so that the reward the agent receives typically remains between -1 and 1, or at least at that magnitude, to improve the stability of agent learning. Therefore, the reward $r_t$ an agent receives at each step is equal to:

\begin{equation}
r_t = 0.01 N_{\text{agents}} R^{\text{sum}}
\end{equation}

Using the raw energy cost as the reward will punish the agent when the demand is high and reward the agent when RES generation is high, neither of which the agent has any control over. Therefore, a baseline $R^{\text{base}}$ is used to normalise the reward for effective learning.  The single-agent algorithms have control over the whole environment so are rewarded on the primary microgrid's energy savings with a $R^{\text{base}}$ of if all of the ESS and MGA had remained idle by recalculating what the reward would be if $x^{\text{LIB}}$, $x^{\text{LIB}}$, $x^{\text{LIB}}$, and $\text{X}^{\text{MGA}}$ were all 0.

However, the multi-agent methods can be more flexible with reward function design. For example, each agent could be rewarded on the entire primary microgrid's savings as this is the common goal the agents are working towards but this could lead to credit assignment problems as the agents will need to assess how much of the shared reward they contributed to. Alternatively, they could be rewarded based on their own individual savings, but that then ignores the policies of the other ESSs and may lead to a lower global reward.

Therefore, both approaches can be combined by rewarding the agent on the grid's savings but by using a baseline of if that one individual agent had remained idle, rather than if all of them together were idle. This follows the principle of marginal contribution in cooperative game theory where agents are rewarded based on their contribution to the global goal \cite{osborne_course_1994}.
\section{Reinforcement Learning Methodology}
\label{sec:RL}
This section will explore the theory behind RL and introduce the specific algorithms used in the simulations later in the paper.

\subsection{Fundamentals}
Each time-step $t$ in RL, the agent observes the current state of the environment $s_t$ from and selects an action $a_t$ from an action-space following a learnt policy. The agent then receives a reward $r_t$ from a reward function and transitions to a new state $s_{t+1}$ following the state transition function. This is shown in Figure \ref{fig:rl}. The goal of the agent is to learn an optimal policy that maximises its total discounted future reward $G$, scaled with a discount factor $\gamma$:

\begin{figure}[!t] 
	\centering
	\includegraphics[width=.45\textwidth]{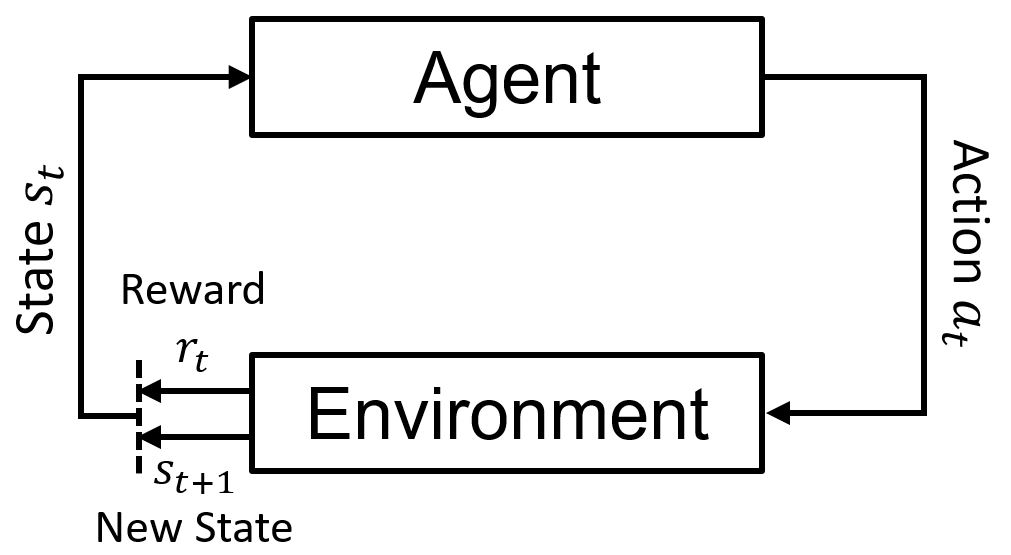}
	\caption{Reinforcement learning control process.}
	\label{fig:rl}
\end{figure}

\begin{equation}
G_t = r_{t}+\gamma r_{t+1}+\gamma^2 r_{t+2}+\ldots = \sum\limits_{k=0}^{\infty}\gamma^{k}r_{t+k}
\end{equation}

The key benefit of RL is that the methods are typically model-free meaning the state transition function and reward function do not need to be known. This makes them incredibly versatile and easy to implement as the agents learn entirely from their own experiences; effectively through trial-and error \cite{sutton_reinforcement_2018}. In contrast, model-based solutions can often perform better due to the ability to plan, but require explicit knowledge about the environment's state and dynamics which may not always be available \cite{sutton_integrated_1990}. 

RL algorithms can be categorised as value function methods, policy gradient methods, or actor-critic methods. Value function methods evaluate performance by assigning value to each state by estimating the state-value $V(s)$ or state-action pair by estimating the action-value $Q(s,a)$. The control problem is then solved recursively using the Bellman equation \cite{bellman_theory_1954}:

\begin{equation}
Q(s_t, a_t) = \mathbb{E} \mathcal{R}(s_t, a_t) + \gamma \mathbb{E}Q(s_{t+1},a_{t+1})
\end{equation}

Policy gradient methods instead directly parameterise the policy without necessarily estimating a value function. The parameters $\theta$ are adjusted to maximise an objective $J(\theta)$ by following the gradient of the policy $\nabla_{\theta}J(\theta)$.

Actor-critic methods will both learn a value function and parameterise the policy, combing both approaches.

\subsection{Algorithms}
\label{section:alg}

The algorithms used in this paper are Deep Deterministic Policy Gradient (DDPG) and two more advanced variants of DDPG, benchmarked against a popular value function method.

\subsubsection{Deep Deterministic Policy Gradient (DDPG)}
DDPG \cite{lillicrap_continuous_2016} is an actor-critic method based on the principles of Q-learning \cite{watkins_learning_1989} in which an actor network $\mu_{\phi}(s)$ with network weights $\phi$ selects the actions the agent takes while a critic network $Q_{\theta}(s,a)$ with weights $\theta$ evaluates agent performance.

Action selection is performed by passing the current state through the actor network. As the policy is deterministic, the agent explores by adding a a noise process $w$ to the actor output:
\begin{equation}
a_t = \mu_{\phi}(s_t) + w
\end{equation}

This noise is typically either Gaussian for uncorrelated noise or the Ornstein–Uhlenbeck process for correlated noise. An alternative method is to instead inject noise into the parameters of the ANNs so that the noise can be adjusted during training \cite{plappert_parameter_2018}. NoisyNet layers \cite{fortunato_noisy_2018} have been used with Deep Q-Networks (DQN) \cite{mnih_human-level_2015} to introduce noise to the value function calculation which the agent can learn to increase or decrease as it visits those states. The same approach is used in this work, and the noise process for the action selection has been removed.

Once the environment has executed the action, the agent stores the transition tuple $\langle s_t,a_t,r_t,s_{t+1},d_t \rangle$ in an experience replay buffer memory, which can then be sampled later during training. This makes the algorithm an off-policy method as the agent is trained from samples collected using a different policy to the current learnt one.

Fixed target actor $\hat{\mu}_{\hat{\phi}}$ and target critic $\hat{Q}_{\hat{\theta}}$ networks are also used to stabilise learning, with weights that are fixed during training and then updated at the end of each step. This is so that the agent is not calculating the target value using the same network that is being trained, which is done to avoid oscillations or divergence in the policy \cite{mnih_human-level_2015}.

The critic and target critic networks evaluate performance by observing both the state and the corresponding action. Following the principles of temporal difference learning \cite{sutton_reinforcement_2018}, target values $y$ are estimated from the target critic estimations of the sampled batch of tuples:
\begin{equation}
y_i = r_i + \gamma \hat{Q}_{\hat{\theta}} \left( s_{i+1},\hat{\mu}_{\hat{\phi}}(s_{i+1}) \right)
\end{equation}

The critic network is updated through gradient descent using the mean-squared error between $y$ and the predicted value from the critic network:
\begin{equation}
\mathcal{L}_i(\theta) = \mathbb{E} \left[ (y_i - Q_{\theta}(s_i,a_i))^2 \right]
\end{equation}

The actor network is then updated through gradient ascent using the deterministic policy gradient \cite{silver_deterministic_2014}:
\begin{equation}
\nabla_{\phi}J_i(\phi) = \mathbb{E} \left[ \nabla_a Q_\theta(s_i,a)|_{a=\mu(s)} \nabla_{\phi}\mu_{\phi}(s_i) \right]
\end{equation}

At the end of each step, the target critic and target actor weights are updated:
\begin{equation}
\begin{aligned}
\hat{\theta} \leftarrow \tau \theta + (1 - \tau) \hat{\theta}\\
\hat{\phi} \leftarrow \tau \phi + (1 - \tau) \hat{\phi}
\end{aligned}
\end{equation}

\begin{figure*}[!t]
\centering
	\begin{subfigure}{.35\textwidth}
		\centering
		\includegraphics[width=\textwidth]{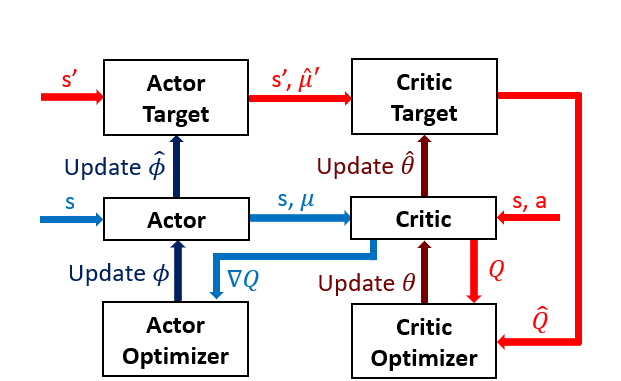}
		\caption{Learning process of DDPG.}
		\label{fig:ddpg}
	\end{subfigure}
	\begin{subfigure}{.48\textwidth}
		\centering
		\includegraphics[width=\textwidth]{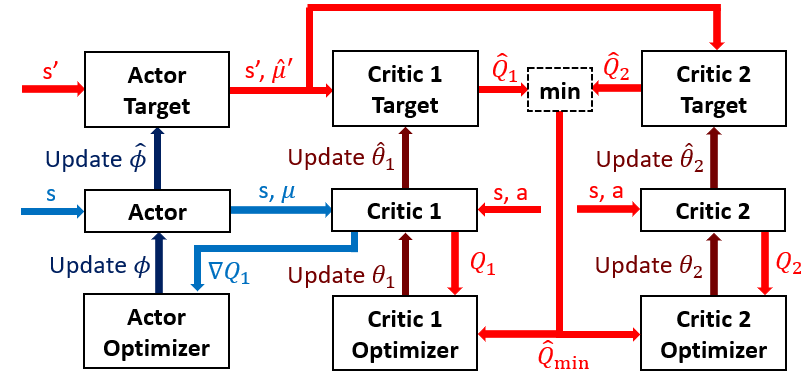}
		\caption{Learning process of TD3.}
		\label{fig:td3}
	\end{subfigure}
	\caption{Learning at each step when provided with a batch of transitions $\langle s_t, a_t, r_t, s_{t+1} \rangle$. The red arrows show the process for training the critic network and the blue arrows for training the actor network.}
	\label{fig:learning}
\end{figure*}

\begin{algorithm}[h] 
\caption{Deep Deterministic Policy Gradient}
\begin{algorithmic}[1]

\STATE Initialise critic $Q_\theta(s,a)$ and actor $\mu_\phi(s)$ arbitrarily
\STATE Initialise target critic $\hat{\theta}$ with $\hat{\theta} \leftarrow \theta$
\STATE Initialise target actor $\hat{\mu}$ with $\hat{\phi}\leftarrow\theta^{v}$
\STATE Initialise replay memory

\FOR{$t=0, T$}
	\STATE Observe state $s_t$
	\STATE Choose action $a_t=\mu(s_t)$ 
	\STATE Execute $a_t$, observe reward $r_t$ and next state $s_{t+1}$
	\STATE Store transition $\langle s_t,a_t,r_t,s_{t+1} \rangle$ in memory
	\STATE Sample minibatch $\langle s_i,a_i,r_i,s_{i+1} \rangle$ from memory
	\STATE Calculate target $y_i = r_i + \gamma \hat{Q}_{\hat{\theta}}(s_{i+1},\hat{\mu}_{\hat{\phi}}(s_{i+1}))$
	\STATE Critic loss $\mathcal{L}_i(\theta) = \mathbb{E} \left[(y_i - Q_{\theta}(s_i,a_i))^2\right]$
	\STATE Perform gradient descent on $\theta$ using $\mathcal{L}(\theta)$
	\STATE Actor loss $\nabla_{\phi}J(\phi) = \mathbb{E}\left[ \nabla_a Q_\theta(s,a)|_{a=\mu(s)} \nabla_{\phi}\mu_{\phi}(s) \right]$
	\STATE Perform gradient descent on $\phi$ using $\nabla_{\phi}J(\phi)$
	\STATE Update critic weights $\hat{\theta} \leftarrow \tau \theta + (1 - \tau) \hat{\theta}$
	\STATE Update actor weights $\hat{\phi} \leftarrow \tau \phi + (1 - \tau) \hat{\phi}$
\ENDFOR

\end{algorithmic}
\label{alg:ddpg}
\end{algorithm}
The pseudocode of DDPG can be found in Algorithm \ref{alg:ddpg}. The learning process during training between the different ANNs for DDPG is visualised as a flowchart in Figure \ref{fig:ddpg}.

\subsubsection{Distributional DDPG (D3PG)}
An issue with using the expectation value for $Q(s,a)$ is that it can over-generalise in environments with a non-deterministic reward function \cite{bellemare_distributional_2017}. D3PG \cite{barth-maron_distributed_2018} removes the expectation step in the Bellman equation and instead estimates the value distribution $Z(s,a)$:

\begin{equation}
Z(s_t, a_t) \stackrel{D}{=} \mathcal{R}(s_t, a_t) + \gamma Z(s_{t+1},a_{t+1})
\end{equation}

A number of atoms are assigned to each action for $N_{\text{actions}} \times N_{\text{atoms}}$ output neurons with the probability distribution of the return $d(s,a)$ calculated using a softmax applied separately across the atoms for each action. The algorithm C51 \cite{bellemare_distributional_2017}, the distributional variant of DQN, uses 51 atoms across each action. The probability mass on each atom is equal to:

\begin{equation}
d_i(s,a) = \frac{\exp(\theta_i(s,a))}{\sum_j \exp(\theta_j(s,a))}
\end{equation}

Each of these probability masses can then be projected onto a support $z$ with a value equally spaced between a minimum return $v_{\text{min}}$ and a maximum return $v_{\text{max}}$:

\begin{equation}
z_i = v_{\text{min}} + (i-1)\frac{v_{\text{max}}-v_{\text{min}}}{N_{\text{atoms}}-1}
\end{equation}

The sum of these probability masses on each support is equal to $Q(s,a)$. The target used for training is the projected target distribution calculated using the categorical algorithm \cite{bellemare_distributional_2017}:

\begin{equation}
(\Phi \hat{\mathcal{T}} Z_{\hat{\theta}}(s,a))_i = \sum^{N_{\text{atoms}}}_{j=0} \left[ 1 - \frac{\vert [\hat{\mathcal{T}}z_j]^{v_{\text{max}}}_{v_{\text{min}}}-z_i \vert}{\Delta z} \right]^1_0 \hat{d}_j(s_{i+1},a^{*})  
\end{equation}

where $\Phi$ is the projection of the distribution onto $z$ and $\mathcal{T}$ is the distributional Bellman operator. As this generates a discrete probability distribution, the Kullback-Leibler divergence between the predicted value and the projected target is used for the loss function:

\begin{equation}
\mathcal{L}_i(\theta)=D_{\text{KL}}\left( \Phi \hat{\mathcal{T}} Z_{\hat{\theta}}(s_i,a_i) \Vert Z_{\theta}(s_i,a_i)\right)
\end{equation}

\subsubsection{Twin Delayed DDPG (TD3)}
A common issue with using functional approximators in RL is they have a tendency to overestimate value estimates which can lead to suboptimal policies as agents tend to select the actions that they have already assumed are more valuable \cite{thrun_issues_1993}. Double DQN \cite{van_hasselt_deep_2016} looked to solve this by removing the maximisation step in the DQN target calculation and selecting the next state's best action using the evaluation network, but this does not work for actor-critic methods such as DDPG as the action selection and value function estimates are separate. Instead, Double Q-Learning \cite{van_hasselt_double_2010} uses two value function estimates so that the next best action and target estimate is performed using different value estimates.

TD3 \cite{fujimoto_addressing_2018} aims to solve the overestimation problem for DDPG by using a similar approach with a second critic and target critic pair. The target value $y_i$ is then calculated using the minimum estimated of the two critic target networks:
\begin{equation}
y_i = r_i + \gamma \min_{i=1,2} \hat{Q}_{\hat{\theta}_i}(s_{i+1},\hat{\mu}_{\hat{\phi}}(s_{i+1}))
\end{equation}

That shared target value $y$ is then used for the loss function of both critic networks:
\begin{equation}
\begin{aligned}
\mathcal{L}_i(\theta_1) = \mathbb{E} \left[ (y_i - Q_{\theta_1}(s_i,a_i))^2 \right]\\
\mathcal{L}_i(\theta_2) = \mathbb{E} \left[ (y_i - Q_{\theta_2}(s_i,a_i))^2 \right]
\end{aligned}
\end{equation}

In contrast to DDPG where both the actor and critic are updated every step, the actor and target networks are updated at a lower frequency to the critic to minimise the error in the policy update. The actor update uses the deterministic policy gradient of only one of critics:
\begin{equation}
\nabla_{\phi}J_i(\phi) = \mathbb{E} \left[ \nabla_a Q_{\theta_1}(s_i,a)|_{a=\mu(s)} \nabla_{\phi}\mu_{\phi}(s_i) \right]
\end{equation}

The learning process of TD3 is visualised as a flowchart in Figure \ref{fig:td3}.

\subsection{Multi-Agent Reinforcement Learning}
In theory, using RL for a MAS should not be much different from a single-agent settings as the states, actions, and reward function can all remain the same. However, there are a number of fundamental challenges that appear when applying standard single-agent RL algorithms in a MAS.

The main problem is that the environment appears non-stationary from the perspective of a single agent as the behaviours of other agents will change during training. This means that an agent may observe exactly the same state as before and perform exactly the same action but receive a completely different reward based on the actions of other agents, which can make learning incredibly difficult. For example, agents in a competitive setting may over-fit their policy based on the behaviours of other agents, whereas agents in a collaborative setting may suffer from credit assignment problems from a shared reward function.

Multi-agent DDPG (MADDPG) \cite{lowe_multi-agent_2020} looks to solve this by using the principle of centralised learning with decentralised execution. Each agent selects its own action based on its own observations during the execution phase, but the architecture of the DDPG critic allows each agent to evaluate performance by observing the actions taken by all other agents.

As each of the agents possesses its own critic network, they can evaluate their own performance using their own reward function. This means that MADDPG can be used in competitive, collaborative, and mixed settings with only minor adjustments from the original DDPG algorithm.
\section{Case Studies}
\label{sec:methods}
\begin{figure*}[!t] 
	\centering
	\includegraphics[width=.98\textwidth]{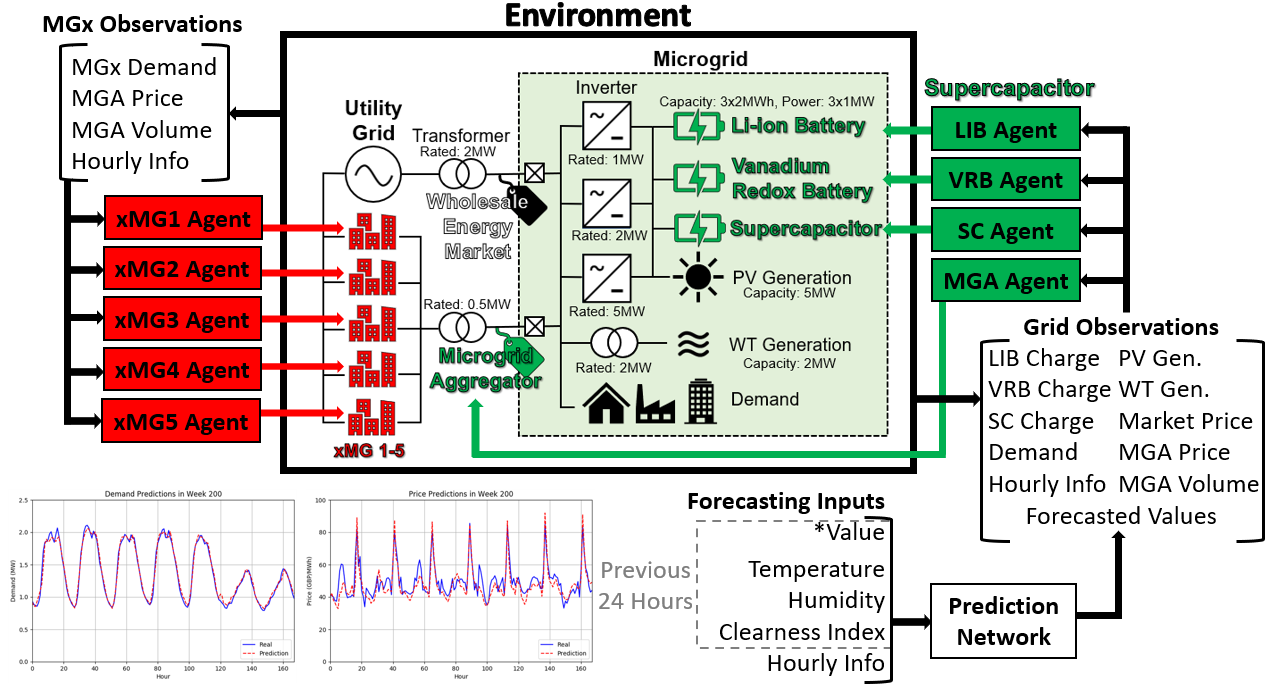}
	\caption{Illustration of how each agent interacts with the environment in the xMG trading case study.}
	\label{fig:project}
\end{figure*}

In the first case study, only the ESS agents are used and their actions control how much each ESS charges or discharges by. In the second, the MGA agent determines the selling volume and reserve price for the next step which all agents observe before the xMG agents place their bids for the amount and price. The control process at each step is found in Figure \ref{fig:project}.

\subsection{Simulation Setup}
The simulations are run across 4 years of data from 2014 up to 2018 with hourly readings for energy consumption, price, and weather parameters. Each time-step represents one hour with each episode representing one week or 168 time-steps. The simulations are run over 200 episodes with the first 100 used for training and the second 100 for evaluation. However, the agents will continue to learn during the evaluation steps so are always learning online. As the environment has no terminal states, the episodes are only relevant for reviewing performance.

Forecasted values are predicted using regression ANNs for the demand, price, and RES generation. In previous work \cite{harrold_data-driven_2022}, this was found to significantly boost agent performance in a similar environment performing energy arbitrage. These forecasts are then provided to the agents as state observations.

Both the forecasting ANNs and algorithms are implemented using \textit{Tensorflow}. For fairness and repeatability, each ANN is set to the same \textit{Tensorflow} and \textit{Numpy} random seed.

\subsection{Benchmarks}
Three benchmarks are used for evaluation in the HESS case study: two alternative multi-agent RL algorithms and a model-based approach. Deterministic optimization methods such as linear programming cannot be used in this environment as the energy price is dependant on whether the agent is buying or selling, as well as the demand and RES. This makes the problem nonlinear and there are no effective methods of solving nonlinear programming problems \cite{boyd_convex_2004}.

No model-based benchmark methods are used for the xMG case study due to the competition between the MGA and the xMG agents. The two alternative multi-agent RL algorithms also can only use discrete action-spaces which give the continuous control of DDPG a tremendous advantage in this particular bidding setting. Therefore, the DDPG algorithms will be benchmarked against each other in the second case study.

\subsubsection{Multi-Agent Deep Q-Networks (MADQN)}
Deep Q-Networks (DQN) \cite{mnih_human-level_2015} uses an ANN to evaluate the action values of different actions, similar to the critic network in DDPG. However, rather than using an actor network for action selection, the agent will always the select the most valuable action at each step unless told otherwise. This makes DQN more robust than actor-critic methods such as DDPG as the algorithms are less sensitive to hyperparameter tuning \cite{sutton_reinforcement_2018}, but the agent can only select from a discrete action-space. Also, as the action selection and evaluation is performed using the same ANN, it is not as easy to apply DQN effectively as DDPG in a MAS.

Foerster et al. \cite{foerster_learning_2016} proposed two approaches for using DQN in a MAS. As MADQN is only being used as a benchmark, the reinforcement inter-agent learning approach is used in which each agent selects an action in succession which is passed onto the next agent as an observation. Therefore, the agents effectively consider all other agents as part of the environment rather than the centralised learning using for MADDPG.

Both MADQN and multi-agent Rainbow (MARainbow) will be used as benchmarks. In previous work \cite{harrold_data-driven_2022}, Rainbow was found to be able to use a larger discrete action-space than regular DQN for ESS control. Therefore, each DQN agent selects from 5 actions whereas the Rainbow agents can select from 9. Rainbow was also shown that its superior learning properties could be more valuable that the continuous control of DDPG.

\subsubsection{Rule-Based Model (RBM)}
The same rule-based approach was used as in previous work \cite{harrold_battery_2020} in which the agent will always look to maximise the utilisation of the RES. This is standard practice for most ESSs looking to maximise RES, but ignores the dynamic energy prices of the wholesale market. The ESSs charge when RES generation exceeds demand and discharge when reversed. As the different ESSs are suited for different timescales of energy storage, the model charges and discharges the SC first, followed by the LIB, followed by the VRB.  
\section{Results and Discussion}
\label{sec:results}
This section will present the results for the two case studies. Discussions will focus on the performance of the agents as well as different behaviours between algorithms, while assessing if the single-agent or multi-agent approaches were better.

The performance of the algorithms is divided into the raw savings, the savings adjusted for the cost of the ESS operation, percentage difference against DDPG, and the percentage of savings lost from the ESS. The second case study also considers the savings made directly from the MGA.

\subsection{Case Study 1: HESS}
\begin{table*}[b]
\caption{Results of Case Study 1, with the best result of each column highlighted.}
\centering
\label{tab:hess_results}
\begin{tabular}{|l|llll|}
\hline
\multirow{2}{*}{Algorithm} & \multicolumn{4}{c|}{Case Study 1} \\
 & Sav.(£1k) & Adj.(£1k) & vs DDPG & ESS Loss \\
\hline
\hline
DDPG & 42.92 & 34.39 & - & 19.87\% \\
D3PG & 48.78 & 40.78 & 18.58\% & 16.40\% \\
TD3 & 46.69 & 39.93 & 16.11\% & 14.48\% \\
\hline
MADDPG & \underline{\textbf{59.58}} & 53.24 & 54.81\% & 10.64\% \\
MAD3PG & 59.50 & \underline{\textbf{53.44}} & \underline{\textbf{55.39\%}} & \underline{\textbf{10.18\%}} \\
MATD3 & 56.19 & 50.06 & 45.57\% & 10.91\% \\
\hline
MADQN & 47.28 & 36.32 & 5.61\% & 23.18\% \\
MARainbow & 52.01 & 43.90 & 27.65\% & 15.59\% \\
RBM & 40.93 & 33.02 & -3.98\% & 19.33 \\
\hline
\end{tabular}
\end{table*}
\begin{figure*}[!t]
\centering
	\begin{subfigure}{.48\textwidth}
		\centering
		\includegraphics[width=\textwidth]{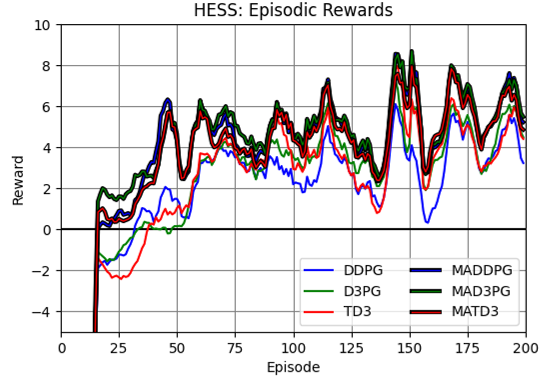}
		\caption{Episodic rewards, averaged over 5 episodes.}
		\label{fig:hess_r}
	\end{subfigure}
	\begin{subfigure}{.48\textwidth}
		\centering
		\includegraphics[width=\textwidth]{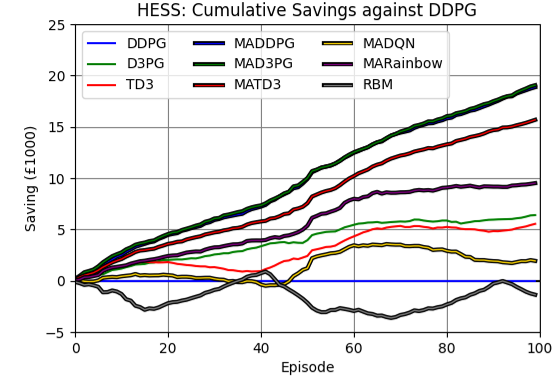}
		\caption{Cumulative savings over each evaluation episode.}
		\label{fig:hess_s}
	\end{subfigure}
	\caption{Rewards and savings of Case Study 1.}
	\label{fig:hess_results}
\end{figure*}

\begin{figure*}[!t] 
	\centering
	\includegraphics[width=.8\textwidth]{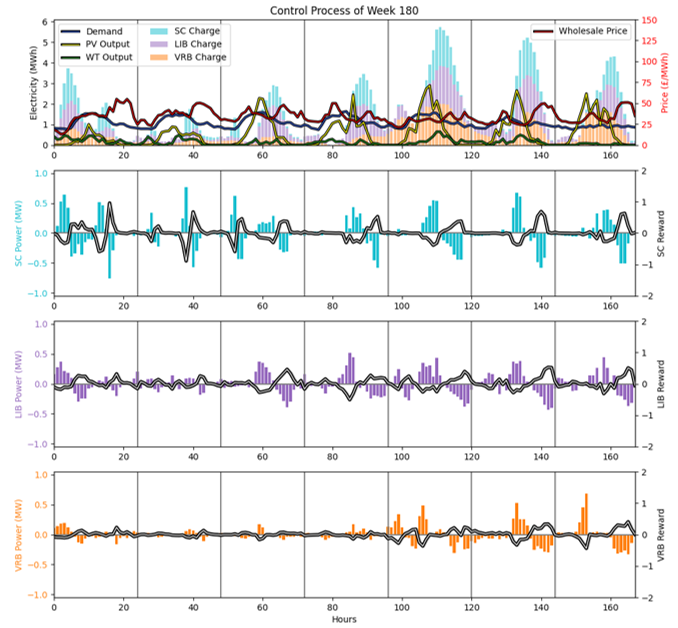}
	\caption{Microgrid control in Episode 180 using MAD3PG.}
	\label{fig:control_hess}
\end{figure*}

This study analysed the use of RL and multi-agent RL for the control of a HESS. The results for this case study can be found in Table \ref{tab:hess_results}, with the smoothed episodic rewards in Figure \ref{fig:hess_r} and adjusted savings over time plotted in Figure \ref{fig:hess_s}. This case study showed that the agents are able to make a significant energy cost savings over two years, even with a low fixed selling tariff by maximising the value of the RES generated.

\subsubsection{Algorithm Performance}
The multi-agent approaches using the improved game theory reward performed overwhelmingly better than the single-agent controllers and the multi-agent methods with shared rewards, with MAD3PG performing 55.4\% better than DDPG. Not only were the multi-agent methods able to use the different ESSs effectively for a lower ESS loss, but were also able to make significantly higher savings before ESS operating cost is considered too. The individual reward functions allow the ESSs operating separately but collaboratively to be much more efficient than having all of them controlled together by a single controller.

\begin{table}[h]
\caption{Adjusted savings of the different approaches.}
\centering
\label{tab:hess_comp}
\begin{tabular}{|l|lll|}
\hline
\multirow{2}{*}{Algorithm} & \multicolumn{3}{c|}{Case Study 1} \\
 & SAS & MAS-S & MAS-MC \\
\hline
DDPG & 34.39 & 40.32 & 53.24 \\
D3PG & 40.78 & 35.27 & \underline{\textbf{53.44}} \\
TD3 & 39.93 & 41.83 & 50.06 \\
\hline
\end{tabular}
\end{table}
However, the multi-agent approaches were only better by such an extent when each agent used the marginal contribution reward. Table \ref{tab:hess_comp} shows the adjusted savings for the different algorithms in both the single-agent system (SAS) and MAS, including the results for if the MAS agents received the same SAS reward rather than the individual MAS rewards using marginal contribution baseline; MAS-S and MAS-MC respectively. All algorithms perform better using the MAS-MC but the effect from SAS to MAS-S is not nearly as consistent, including a notable drop in performance for D3PG. This means the benefit in using a MAS for this environment comes from being able to utilise better reward function design, rather than simply from the distributed control.

MADQN performs fairly well but is clearly limited by the small discrete action-space. Despite this, this method shows the value of using a multi-agent approach as it is still able to outperform DDPG and TD3 with the limitation. MARainbow performs better with the larger action-space but still cannot reach the level of the multi-agent DDPG-based methods as the reinforced inter-agent learning approach appears to be inferior to centralised learning.

RBM performs well but is completely passive on a number of weeks where RES never exceeds demand. This shows the RL agents are able to more effectively utilise the RES, but can also able to respond to the price signals by charging the ESS types when prices are lower in the morning to use later when the prices rise.

\subsubsection{Agent Behaviour}
Most of the agents generally behave in a similar way. The ESSs will only really be used at times when there is a large amount of RES to store with the agents learning that any PV generation at its peak around noon is best to store so it can be used later when prices are higher. There is also some charging at the start of the day when prices are lower, but this is less significant. Due to the battery cycling costs, agents very rarely charge or discharge at maximum power but instead look to charge and discharge over longer periods of time.

Almost every agent uses the LIB the most as its well-rounded properties make it ideal for storing over short, medium, and long-term. The D3PG, MADDPG, MATD3, MADQN, and MARainbow agents then use the SC the second most whereas for DDPG and TD3 it is the VRB. A clear difference between the single controller versus multiple agent methods is that the multi-agent algorithms showed a clear change in the behaviour for the different ESS types, whereas it was much less pronounced in the single agent algorithms.

There are a couple of cases of unique behaviour between algorithms though. Most notably, the MAD3PG agent uses the SC the most aggressively, followed by the LIB and then the VRB which is interesting as its adjusted savings are the best of any other algorithm. The control process of the MAD3PG agent for week 180 is shown in Figure \ref{fig:control_hess}, selected as the episode is towards the end of the simulation but has higher RES generation than later episodes.

Other examples of notable behaviours include D3PG which would look to store energy over greater lengths of time. However, as the demand and price peak cycle with a peak in the evening every day, it is more efficient to use as much stored energy as possible at the peak price times and end the day with minimal charge. Another example is the MATD3 VRB agent which stays almost completely idle all of the time. The RBM behaves as it is instructed, but the resulting issue is that it is completely inactive if RES does not exceed demand whereas the other agents still learn to charge when there are low prices in the morning regardless of RES.

The multi-agent algorithms performed better as the agents could learn a policy for each type of ESS, rather than a single policy for controlling every ESS. Generally, the single agent algorithms would operate each ESS as if they had the same properties because it is more difficult to differentiate the reward punishments for the multiple ESSs when there is only a single reward function. Therefore, the multi-agent approach should be considered superior as it allows the learning of better individual policies for each ESS.

\subsection{Case Study 2: xMG Trading}
\begin{table*}[b]
\caption{Results of Case Study 1, with the best result of each column highlighted.}
\centering
\label{tab:xmg_results}
\begin{tabular}{|l|llllll|}
\hline
\multirow{2}{*}{Algorithm} & \multicolumn{6}{c|}{Case Study 2} \\
 & Sav.(£1k) & Adj.(£1k) & vs DDPG & ESS Loss & MGA  & MGA\% \\
\hline
\hline
DDPG & 151.38 & 146.46 & - & 3.25\% & 129.17 & 88.19\% \\
D3PG & 158.35 & 156.75 & 7.03\% & \underline{\textbf{1.01\%}} & 154.39 & \underline{\textbf{98.49\%}} \\
TD3 & 174.04 & 170.63 & 16.50\% & 1.96\% & \underline{\textbf{157.68}} & 92.41\% \\
\hline
MADDPG & 170.36 & 164.86 & 12.56\% & 3.23\% & 124.72 & 75.65\% \\
MAD3PG & 187.01 & 180.17 & 23.02\% & 3.66\% & 135.84 & 75.40\% \\
MATD3 & \underline{\textbf{190.54}} & \underline{\textbf{186.00}} & \underline{\textbf{27.00\%}} & 2.38\% & 144.68 & 77.78\% \\
\hline
\end{tabular}
\end{table*}
\begin{figure*}[!t]
\centering
	\begin{subfigure}{.48\textwidth}
		\centering
		\includegraphics[width=\textwidth]{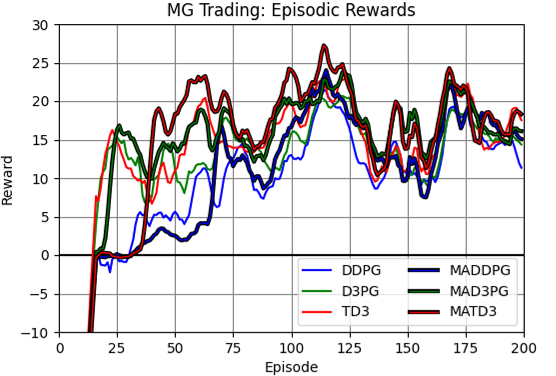}
		\caption{Episodic rewards, averaged over 5 episodes.}
		\label{fig:mg_r}
	\end{subfigure}
	\begin{subfigure}{.48\textwidth}
		\centering
		\includegraphics[width=\textwidth]{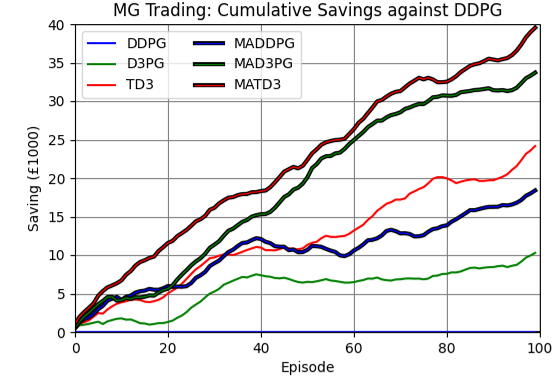}
		\caption{Cumulative savings over each evaluation episode.}
		\label{fig:mg_save}
	\end{subfigure}
	\caption{Rewards and savings of Case Study 2.}
	\label{fig:mg_results}
\end{figure*}

\begin{figure*}[!t] 
	\centering
	\includegraphics[width=.8\textwidth]{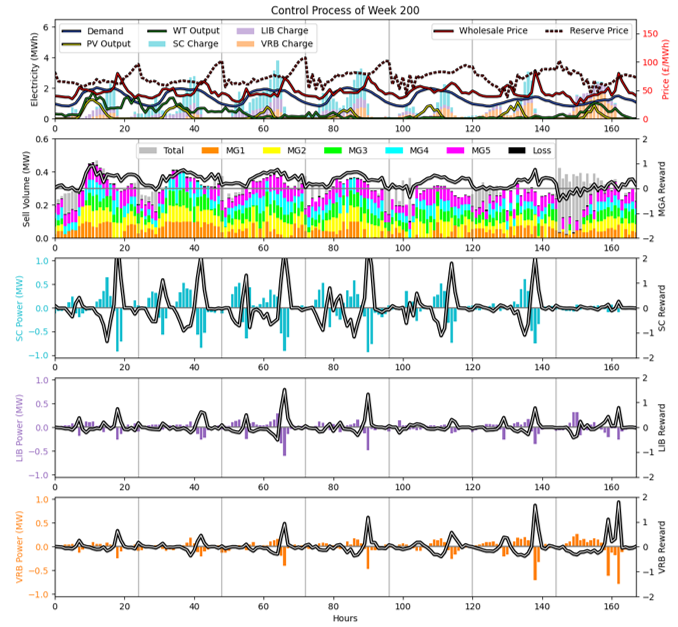}
	\caption{Microgrid control in Episode 200 using MATD3.}
	\label{fig:control_trade}
\end{figure*}

\begin{figure*}[!t] 
	\centering
	\includegraphics[width=.8\textwidth]{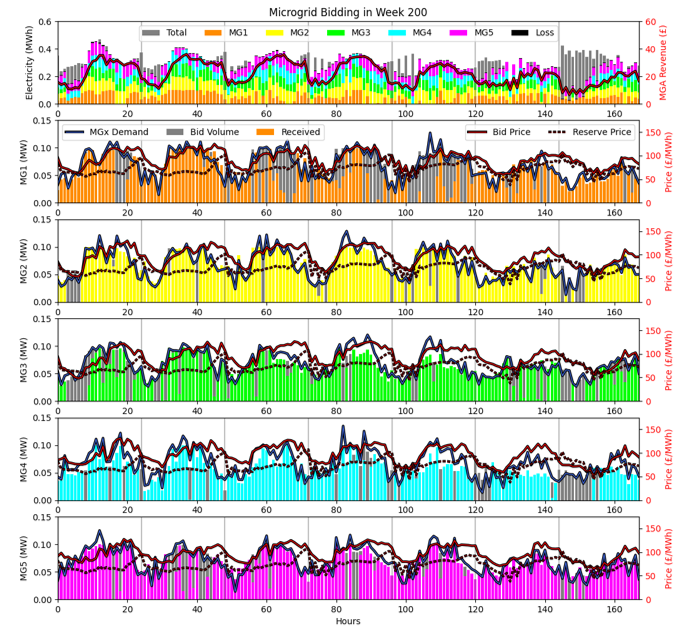}
	\caption{MGA and xMG trading in Episode 200.}
	\label{fig:control_mgx}
\end{figure*}

The results for this case study can be found in Table \ref{tab:xmg_results}, with the smoothed episodic rewards in Figure \ref{fig:mg_r} and adjusted savings over time plotted in Figure \ref{fig:mg_save}. When compared to the results of Case Study 1, the tables shows very clearly that being able to sell energy on the grid's own terms greatly increases savings over simply trying to improve its utilisation. The multi-agent methods once again perform better than the single controller, however the margin difference is significantly smaller.

\subsubsection{Algorithm Performance}
The single-agent methods have a lower adjusted energy cost savings than the multi-agent methods, but interestingly D3PG and TD3 returned a lower ESS loss percentage and a higher MGA profit than any of the multi-agent methods. A key difference between the approaches is that single global controllers try to exploit MGA trading significantly more than the multi-agent approach. D3PG again shows unique behaviour in that it almost entirely ignores the ESSs and instead makes 98.49\% of its savings through MGA trading, significantly higher than the 77.78\% of MATD3. Although interesting that D3PG and TD3 are able to make so much money through trading, this could be a cause for concern as the agents appear to have little interest in exploring how the HESS can be used in conjunction with trading for even higher savings.

\begin{table}[h]
\caption{Adjusted savings of the different approaches.}
\centering
\label{tab:xmg_comp}
\begin{tabular}{|l|lll|}
\hline
\multirow{2}{*}{Algorithm} & \multicolumn{3}{c|}{Case Study 2} \\
 & SAS & MAS-S & MAS-MC \\
\hline
DDPG & 146.46 & 128.89 & 164.86 \\
D3PG & 156.75 & 154.56 & 180.17 \\
TD3 & 170.63 & 182.21 & \underline{\textbf{186.00}} \\
\hline
\end{tabular}
\end{table}
This case study also shows the value of using the marginal contribution reward for the multi-agent methods, shown in Table \ref{tab:xmg_comp}. However, MATD3 using the MAS-S reward is still able to perform remarkably well with a noteworthy improvement over the single agent and only slightly worse than with the MAS-MC reward.

In fact, MATD3 performed the best out of all algorithms with a 27.00\% improvement over DDPG. Interestingly, TD3 also performed very well relative to the other single agent methods and was also able to outperform MADDPG. This suggests that the value overestimation correction from TD3 is particularly useful in this case study. This is likely because, even though the agents are aware that they will receive a much higher reward through trading than through effective RES utilisation, the TD3 agents do not fixate entirely on trading and develop a suboptimal policy where they but instead recognise that effectively using the HESS to effectively utilise RES can enhance performance as well.

\subsubsection{Agent Behaviour}
For the ESSs, SC is used more aggressively for short-term storage, LIB is used for medium-term storage, and the VRB has varied usage between used not at all to usage at the same level as the LIB. In general agents are less aggressive with RES utilisation as any excess RES can be sold rather than just stored, but also the multi-agent methods are more aggressive with the ESSs even when there is little RES generation. Some single-agent methods appear to even completely neglect the ESS operation for large parts of the simulation.

Initially, all agents take random actions for 1000 steps but begin learning after 500. Afterwards, the first observable behaviour comes from the xMGs which immediately learn that buying energy from the MGA is cheaper than buying from the utility grid so attempt to buy as much as possible. However, the MGA agent would learn one of two early behaviours: 

\begin{enumerate}
\item Sell as much as possible at the minimum reserve price, such as MAD3PG and TD3. This leads to a high reward initially but the xMGs quickly learn to exploit this and lower their bid prices until the MGA ends up selling at lower than the wholesale price and losing money.

\item Sell nothing at the maximum reserve price, such as DDPG and MATD3. The MGA will eventually unlearn this behaviour, however the superior learning properties of TD3 over regular DDPG are clear as the MATD3 agent unlearns this significantly quicker than the DDPG and MADDPG agents.
\end{enumerate}

After about 100 episodes, the agents reach an equilibrium and behave in a similar way but at slightly different levels of efficiency. Although the MGA could always sell an amount of energy that would meet the demands of all of the xMGs together, it will instead typically restrict the volume it sells to force the xMGs to compete with each other. An example of the trading is shown in Figure \ref{fig:control_mgx}.

As in the previous case study, the multi-agent approaches perform better than the single agent controllers. The separate reward functions for the different ESSs allows for each to learn their own policy suited to their characteristics while also allowing those agents to not become fixated on the high MGA rewards that the single agents fell short to.

This case study exhibits the benefit of being able to sell energy on the primary microgrid's terms, rather than only selling back to the utility grid. Trading is beneficial to all parties in that the primary microgrid is able to generate additional revenue, the xMGs are able to reduce their own energy bills, and inturn minimises the load on the utility grid. 

\subsection{Improvements and Further Work}
One improvement would be expanding the functionality of the xMGs. This would include having their own ESSs and RES but also their own aggregator agents that would be able to trade with the other xMGs. This would take away much of the MGA's influence on the xMG trading phase and could allow the xMGs to develop their own unique behaviours.

A flaw in the environment is that wholesale energy prices are used from a day-ahead market, rather than a real-time market. To be more realistic, the primary microgrid should decide the volume of energy it wishes to import from the utility grid at least 24 hours ahead. A more accurate feed-in tariff model based which varies based on the amount and type of energy sold could also be considered. However, the agents already learn to minimise the amount of energy sold back to the utility grid so would likely have a negligible impact on results or their behaviour. 

Another improvement is forecasting further into the future for more efficient long-term storage. In most cases across both case studies, the agents tend to avoid using the VRB due to the poor cycling efficiency, but it would be interesting to see if the agents would change this behaviour if there were more forecasts provided so that the long-term storage properties of the VRB could be exploited to more effect.

Further work could look into other methods of communication for multi-agent RL. For example, differential inter-agent learning \cite{foerster_learning_2016} allows the principle of centralised learning to be applied to DQN and Rainbow, rather than the current approach where the other agents are treated as part of the environment. There could also be other types of agents in the grid, such as a specific agent for purchasing energy from a day-ahead market or an agent controlling a distributed generator powered by diesel or biofuel.
\section{Conclusion}
\label{sec:conclusion}
In this paper, the use of multi-agent reinforcement learning was presented for the control of a HESS in a microgrid to increase renewable energy utilisation, reduce energy bills, and trade energy to xMGs. Specifically, the principle of centralised learning and decentralised execution with MADDPG is explored where agents can learn their own policies but evaluate their performance by considering the policies of every agent in the MAS, as well as more advanced variants of DDPG in D3PG and TD3.

The research found that the multi-agent approaches where each agent, with its their own reward function, controls an ESS performed better than a single global agent controlling the entire network. The separate reward functions for each component allowed the agents to more effectively evaluate their individual contribution to the shared goal, following the principle of marginal contribution from game theory. It was also found that selling energy to xMGs through the aggregator was significantly more profitable than simply maximising RES utilisation or selling back to the utility grid at a fixed feed-in tariff. Trading energy between the agents was beneficial to the primary microgrid through increased revenue, allowed the xMGs to reduce their own energy bills, and reduced the load on the main utility grid.

\section*{Acknowledgements}
This work was supported by the Smart Energy Network Demonstrator project (grant ref. 32R16P00706) funded by ERDF and BEIS. This work is also supported by the EPSRC EnergyREV project (EP/S031863/1) and the State Key Laboratory of Alternate Electrical Power System with Renewable Energy Sources (Grant No. LAPS20012).

\bibliographystyle{elsarticle-num}
\bibliography{MAS_Project}

\end{document}